# DECISION SUPPORT FOR E-GOVERNANCE: A TEXT MINING APPROACH

G. Koteswara Rao[1] and Shubhamoy Dey [2]

[1]Information Systems, Indian Institute of Management, Indore, M.P, INDIA
gkrao@iimidr.ac.in
[2]Information Systems, Indian Institute of Management, Indore, M.P, INDIA
shubhamoy@iimidr.ac.in

## ABSTRACT

*Information and communication technology has the capability to improve the process by which governments involve citizens in formulating public policy and public projects. Even though much of government regulations may now be in digital form (and often available online), due to their complexity and diversity, identifying the ones relevant to a particular context is a non-trivial task. Similarly, with the advent of a number of electronic online forums, social networking sites and blogs, the opportunity of gathering citizens' petitions and stakeholders' views on government policy and proposals has increased greatly, but the volume and the complexity of analyzing unstructured data makes this difficult. On the other hand, text mining has come a long way from simple keyword search, and matured into a discipline capable of dealing with much more complex tasks. In this paper we discuss how text-mining techniques can help in retrieval of information and relationships from textual data sources, thereby assisting policy makers in discovering associations between policies and citizens' opinions expressed in electronic public forums and blogs etc. We also present here, an integrated text mining based architecture for e-governance decision support along with a discussion on the Indian scenario.*

## KEYWORDS

*Text mining techniques, e- governance, public policy, public opinion, decision support systems*

## 1. INTRODUCTION

Data mining was conceptualized in the 1990s as a means of addressing the problem of analyzing the vast repositories of data that are available to mankind, and being added to continuously. Considering the fact that most data (over 80%) is stored as text, text mining has even higher potential [2]. Text mining is a relatively new interdisciplinary field that brings together concepts from statistics, machine learning, information retrieval, data mining, linguistics and natural language processing. It is said to be the discovery by computer of new, previously unknown information by automatically extracting information from different written resources [3]. Text mining is different from mere text search or web search where the objective is to discard irrelevant material to identify what the user is looking for. Essentially, in the context of text search, the user knows what he / she is looking for (in the form of keywords etc.), and the (written) material already exists. In text mining one of the key elements is that the aim is to discover unknown information by linking together existing text data to form new facts or hypotheses. Thus, in many ways text mining is similar to data mining, and indeed regarded by some as an extension of the same. The main point of departure from the parent discipline of data mining is in the type of data that needs to be analyzed. Whereas data mining deals with mostly numeric structured data, text, the theme of text mining, is regarded as 'unstructured' data. Though, the task of text mining based DSS would seem to be more challenging than that of mining of structured data, the existence of vast amounts of information in electronically available text has led to intense research in text mining techniques, and many of the challenges have been overcome.

The greatest potential of applications of text mining is in the areas where large quantities of textual data is generated or collected in the course of transactions. For example industries like publishing, legal, healthcare and pharmaceutical research, and areas like customer complaints (or feedback) handling and marketing focus group programs would be the best areas of application of text mining. Innovative applications in the contexts of personalization in B2C e-commerce, competitive intelligence, customer satisfaction analysis and e-mail filtering are discussed in numerous articles [4-7]. Not surprisingly, text mining has been successfully applied for the purpose of easing the tedium of content analysis and literature survey in research work [8],[9].

Decision support systems (DSS) help leaders and managers make decisions in situations that are unique, rapidly changing, and not easily specified in advance [01]. Text Mining based DSS (TMbDSS) integrate unstructured textual data with predictive analytics to provide an environment for arriving at well-informed citizen-centric decisions in the context of e-governance.

## 2. TEXT MINING BASED DECISION SUPPORT (TMbSS): TECHNIQUES AND ARCHITECTURE FOR E-GOVERNANCE

The technologies used in TM include: information retrieval (IR), information extraction (IE), topic tracking, summarization, categorization, concept linkage, information visualization, and question answering. The most widely used text mining techniques [10] are discussed briefly below to enable better understanding of their application in the field of e-governance, citizen participation and e-democracy.

1. Information extraction: Information extraction algorithms identify key phrases and relationships within text. This is done by looking for predefined sequences in text, using a process called 'pattern matching'. The algorithms infer the relationships between all the identified sequences to provide the user with meaningful insight. This technology can be very useful when dealing with large volumes of text.
2. Categorization: Categorization involves identifying the main themes of a document by placing the document into a pre-defined set of topics. It does not attempt to process the actual information as information extraction does. Categorization only counts words that appear in the text and, from the counts, identifies the main topics that the document covers. Categorization often relies on a thesaurus for which topics are predefined, and relationships are identified by looking for broader terms, narrower terms, synonyms, and related terms.
3. Clustering: Clustering is a technique used to group similar documents, but it differs from categorization in that documents are clustered based on similarity to each other instead of through the use of predefined topics. A basic clustering algorithm creates a vector of topics for each document and measures how well the document fits into each cluster.
4. Topic tracking: A topic tracking system works by keeping user profiles and, based on the documents the user views, predicts other documents of interest to the user. Some of the better text mining tools let users select particular categories of interest, and can even automatically infer the user's interests based on his/her reading history and click-through information.
5. Summarization: Text summarization is immensely helpful for trying to figure out whether or not a lengthy document meets the user's needs and is worth reading for further information. The key to summarization is to reduce the length and detail of a document while retaining its main points and overall meaning.
6. Question answering: Another application area of text mining is answering of question answering, which deals with how to find the best answer to a given question. Question answering can utilize more than one text mining techniques.

7. Association detection: In Association Rules, the focus is on studying the relationships and implications among topics, or descriptive concepts, which are used to characterize a set of related text. The goal is discover important association rules within a corpus such that the presence of a set of topics in an article implies the presence of another topic.

As per Rao et al [11], Text mining techniques, though relatively new, are considered mature enough to be incorporated into almost all commercial data mining software packages. The features of some popular data mining software that have text mining modules are summarized in their paper. They have observed that text mining has made a transition from the domain of research to that of robust industrial strength technology, and can be used in mission critical applications like e-governance.

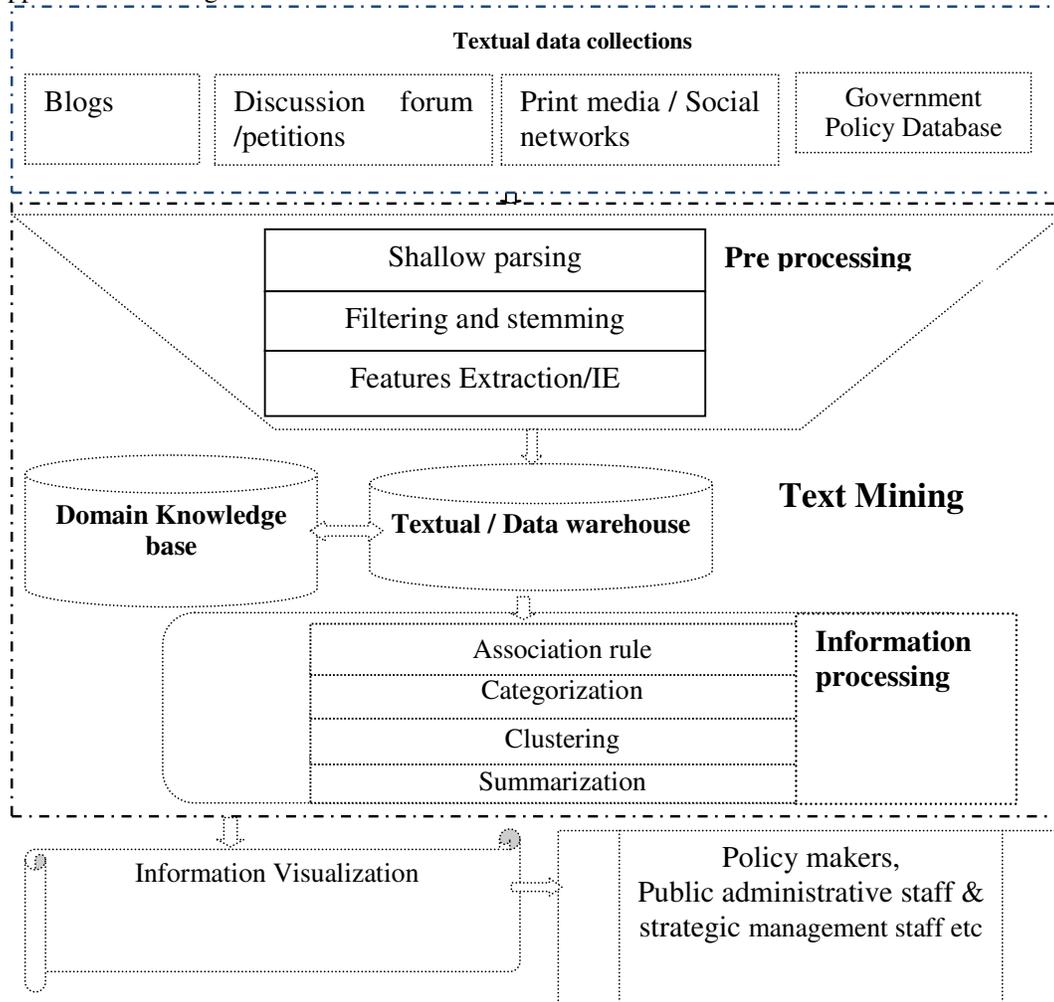

Figure. 1. Text mining based decision support system for e-government: technical architecture

Apart from the commercial text mining packages, a number of open source software packages are also available. Most of these packages being available free or at low cost could be useful for pilot projects, enabling early adopters to move forward without making heavy financial commitments. The following table (Table-1) lists a few of the open source text mining packages. Figure 1 illustrates the basic technical architecture for Text-Mining based DSS for e-governance. Yue Dai et al, have proposed a similar architecture for a system for competitive intelligence in a decision support system model called MinEDec (Mining Environment for Decisions), which is supported by text-mining technologies [12].

**Table 1.**

| Open source | Description |
|---|---|
| **Carrot2** http://project.carrot2.org | Carrot2 is an Open Source Search Results Clustering Engine. It can automatically organize small collections of documents, e.g. search results, into thematic categories.Carrot2 offers ready-to-use components for fetching search results from various sources including GoogleAPI, Bing API, eTools Meta Search, Lucene, SOLR, Google Desktop and more. |
| **GATE** http://gate.ac.uk | Open source software capable of solving almost any text processing problem. In active use for all sorts of language processing tasks and applications, including: voice of the customer; cancer research; drug research; decision support; recruitment; web mining; information extraction; semantic annotation. Many educational institutes have included GATE in their text Mining courses. |
| **Natural Language Toolkit (NLTK)** http://www.nltk.org | A suite of libraries and programs for symbolic and statistical natural language processing (NLP) for the Python programming language. NLTK comes with many corpora, toy grammars, trained models, etc. NLTK is suited to courses in many areas, including natural language processing, computational linguistics, empirical linguistics, cognitive science, artificial intelligence, information retrieval, and machine learning. |
| **RapidMiner** http://rapid-i.com/content/view/181/190 | Formally called as YALE Yet Another Learning Environment), is an environment for machine learning, data mining, text mining, predictive analytics, and business analytics. The Text mining plugin contains tasks specially designed to assist on the preparation of text documents for mining tasks, such as tokenization, stop word removal and stemming. RapidMiner plugins are Java libraries that need to be added to the lib\plugins subdirectory under the installation location. |
| **Unstructured Information Management Architecture (UIMA)** http://uima.apache.org | It was originally developed by IBM. It is an open, industrial-strength, and scalable and extensible platform for creating, integrating and deploying unstructured information management solutions from combinations of semantic analysis and search components. UIMA's goal is to provide a common foundation for industry and academia to collaborate and accelerate the world-wide development of technologies critical for discovering the vital knowledge present in the fastest growing sources of information today. |
| **tm:Text Mining Package** http://cran.r-project.org/web/packages/tm/index.html | It offers functionality for managing text documents abstracts the process of document manipulation and eases the usage of heterogenous text formats in R. The package has integrated database backend support to minimize memory demands. An advanced meta data management is implemented for collections of text documents to alleviate the usage of large and with meta data enriched document sets. |

To implement any intelligent system the primary step is the selection of required sources, which in our case are, the government policy database, the citizens' complaints from relevant web portals, the online discussion forums, to allow citizens' to discuss about prestigious government projects and last but not the least the social network/media, which have gained immense popularity in modern times as one can extract the political data from social network /media to understand the stakeholders opinions. As we are talking about the unstructured information from multiple sources and in different formats (pdf, doc, docs, xml,jpg,html etc.) we need to use parsing system to transform the documents into the format, which has the capability to handle unstructured/semi-structured data. Next task is the information (keyword/ features) retrieval; it includes tokenization, filtering, stemming, indexing and refinement. However, in some cases traditional keyword extraction techniques may not be able to support, we would then need to implement other techniques to extract features that include generic features, domain-specific features and concepts extraction and then refine the regulation database. After the features and information have been stored in the textual/data warehouse, association rule analysis, clustering, categorizing, and summarization can be used to process them into meaningful information.

## 3. TEXT MINING APPLICATIONS IN E-GOVERNANCE

The transformation from conventional government services to E-government services heralds a new era in public services. E-government services can replace the government's traditional services with services of better quantity, quality and reach, and increase citizen satisfaction, using Information and Communication Technology (ICT). E-governance aims to make the interactions between government and citizens (G2C), government and business enterprise (G2B) and inter-government department dealing (G2G) friendly, convenient transparent and less expensive [13]. A growing amount of informative text regarding government decisions, directives, rules and regulations are now distributed on the web using a variety of portals, so that citizens can browse and peruse them. This assumes, however, that the information seekers are capable of untangling the massive volume and complexity of the legally worded documents [14]. Government regulations are voluminous, heavily cross-referenced and often ambiguous. Government information is in unstructured / semi-structured form, the sources are multiple (government regulations comes from national, state and local governments) and the formats are different – creating serious impediment to their searching, understanding and use by common citizens.

In the G2G arena, the government departments are in an even greater need of a system that is able to provide information retrieval, data exchange, metadata homogeneity, and proper information dissemination across the administrative channels of national, regional / state, and local governments [15]. The increasing demand for and complexity of government regulations on various aspects of economic social and political life, calls for advanced knowledge-based framework for information gathering, flow and distribution. For example, if policy makers intend to establish a new act, they need to know the acts related to the same topic that have been established before, and whether the content of the new act conflicts with or has already been included in existing acts [16]. Also, regulations are frequently updated by government departments to reflect environmental changes and changes in policies. Tools that can detect ambiguity, inconsistency and contradiction are needed [16] because the regulations, amended provisions, legal precedence and interpretive guidelines together create a massive volume of semi-structured documents with potentially similar content but possible differences in format, terminology and context. Information infrastructures that can consolidate, compare and contrast different regulatory documents will greatly enhance and aid the understanding of existing regulations and promulgation of new ones.

Government regulations should ideally be retrievable and understandable with ease by legal practitioners, policy makers as well as general public /citizens. Despite many attempts, it is recognized that e-government services are yet to render the desired pro-citizen services and are mostly targeted towards internal efficiency [13]. Kwon et al [22], have proposed a system that helps rule makers understand and respond to the public comments, before finalizing proposed regulations [22]. These public comments are opinion-oriented arguments about the regulations.

The facility of identification and classification of main subject of the claims / opinions provided by the tool helps rule-writers preview and summarize the comments [22]. The proposed solution identifies conclusive sentences showing the author's attitude towards the main topic and classifies them to polar classes [22]. The researchers have applied a supervised machine learning method to identify claims using sophisticated lexical and structural features and to classify them by the attitude to the topic: in support of, opposed to, and proposing a new idea [20].

## 4. INTEGRATING CITIZENS VOICE WITH E-GOVERNANCE THROUGH TMbDSS

It is widely acknowledged that democracy requires well-informed citizens. Information creates trust and is the mechanism for ensuring that politicians serve the electorate. Democracy if effective when there is smooth flow of information between citizens and government [17]. E-governance in its present form has furthered this concept to a certain extent. However, the character of e-governance is mainly one-way flow of information – from the government to the citizens, and authentic citizen participation is absent. With the integration of citizens' participation in the entire process of governance with the help of Information and Communication Technology e-governance evolves into E-democracy and Citizen Participation in policy making can secure democracy, as it generates a continuous flow of information between citizens and the government, helping them in the decision-making process and the citizens can assume a more active role in society, exercising their opinion power with ease and agility [18].

In the usual form of democracy, the general election is the most important citizen participation process. It is significant because it formulates the country's transfer of power from one civilian government to another. Since, elections are intermittent, it is important to have a system in place that has the capability to track public opinion on a more or less continuous basis, and encourage involvement and participation from the electorate on matters of public importance [17]. It is quite possible for citizens' to have different opinions on government proposals. Government can use the online discussion forums and encourage citizens' to discuss on public projects. Once the discussions phase is opened and finished its output are needs to be analyzed so that the underlying trends and preferences of citizens can be incorporated into the decision-making process of the pertinent administrative department [19]. Capturing citizens' opinions through electronic participation / discussion media can be more reliable than traditional methods based on opinions polls and help avoid false opinion declaration. This also drastically changes the methods of surveying citizens' opinion trends as well as the accuracy of the evaluation of their opinions. It reduces the cost, increases reach, and provides almost real time information. Potentially, arguments that led to significant opinion shifts can be detected. However, the volume and the complexity of analyzing unstructured data make this far from straight forward. Text mining can process unstructured data leading to greater understanding of the text in the context of others on the same topic. This is especially important when dealing with expressed public opinion, where the arguments for and against particular positions are important to identify and gauge, but is immensely difficult to extract due their storage in natural language format [20].

Cardeñosa [19] proposes a system, which has the capability to process the messages posted by citizens' on e-message boards, e-mails and open debate threads etc. It collects the messages from online forums, classifies them, identifies the supporting expressions, and extracts the common features and regularities. The system uses association rule mining technique to identify the trend between the citizens' opinions. These rules form the intelligent core of the system. The future refinements and extensions of the system are in the direction of building a more accurate voting pattern prediction system. Fatudimu [21] has developed a system to process the unstructured data from newspaper articles to understand the stakeholders' opinions on elections. The system proposed applies text-mining techniques on the information collected through newspapers and applies natural language processing (NLP) and association rule mining to extract knowledge and understand the citizens' voice on election issues.

Luehrs et al, have discussed about Online Delphi Survey module and how it can be used to conduct online surveys and enhance citizens' participation in public issues. And he also discussed about how citizens' discussions on public issues can be analyzed qualitatively and categorize by using text-mining algorithms based on standard Bayesian inference methods. The proposed solution can be used to extract the 'concepts' or main ideas out of a free text and to search for 'similar texts' based on comparison of these concepts [23].

Scott et al, opine that Social networking sites can be viewed as a new type of online public sphere. They have discussed a system that they have implemented to examine the linkage patterns of citizens' who posted links on Facebook "walls" of Barack Obama, Hillary Clinton, and John McCain over two years prior to the 2008 U.S. elections [24]. Web logging (blogging) and its social impact have recently attracted considerable public and scientific interest. Tae Yano et al have collected blog posts and comments from 40 blog sites focusing on American politics during the period November 2007 to October 2008, contemporaneous with the presidential elections. They have concluded that predicting political discourse behaviour is challenging, in part because of considerable variation in user behaviour across different blog sites. Their results show that using topic modelling; one can begin to make reasonable predictions as well as qualitative discoveries from the language used in blogs [25].

Muhlberger et. al, have implemented an Interactive Question Answering (QA), Dialogue Analysis, and Summarization into a viable learning and discussion facilitation agent called the Discussion Facilitation Agent (DiFA), which will try to keep users(citizens) informed, on the fly, about changes and developments in the deliberation content, and summarize key arguments at the conclusion. [26]. A few other similar systems have been developed by other researchers like Pérez, et. al [27] and Maciel and Garcia [28].These systems though somewhat futuristic and still in the process of being researched, demonstrate that the concept of participation of citizens' in democratic processes through electronic media is an achievable one. It is also evident from the way these systems work, that text mining capability is the cornerstone of the move towards e-democracy systems.

Figure-2 depicts a 'Participation System' for gathering, analysis and addressing citizens' concerns regarding existing / proposed government policies / laws. In the figure, the central repository of documents (mostly in unstructured form) has been labelled 'Proposed Govt policies/Govt policies. The citizens are encouraged to record their reactions through the 'public forums / feedback'. Government can also collect data corpus from Social networks. Print/Digital Media contains data in the form of 'Public dialogue and stakeholders opinions. Each of these three corpuses contains huge amount of unstructured/semi structured Data. Knowledge/ insights extracted from these databases can be used in forming new regulation/policies, understanding citizens' opinions and answering their concerns. The main users of the system are Public Administrative officers (PA Officers), Moderators and Decision makers. It helps in the formulation of new policies, budget analysis, understanding the stakeholders' opinion on national level projects and regulations with the help of text mining tools. Government agencies can better understand social behaviour and demands, through analyzing citizens' behaviour  patterns, information  extracted from this can be used to  provide citizen centric solution and maintain a closer relationship between government and citizens and enhance the citizens' satisfaction on govt services.

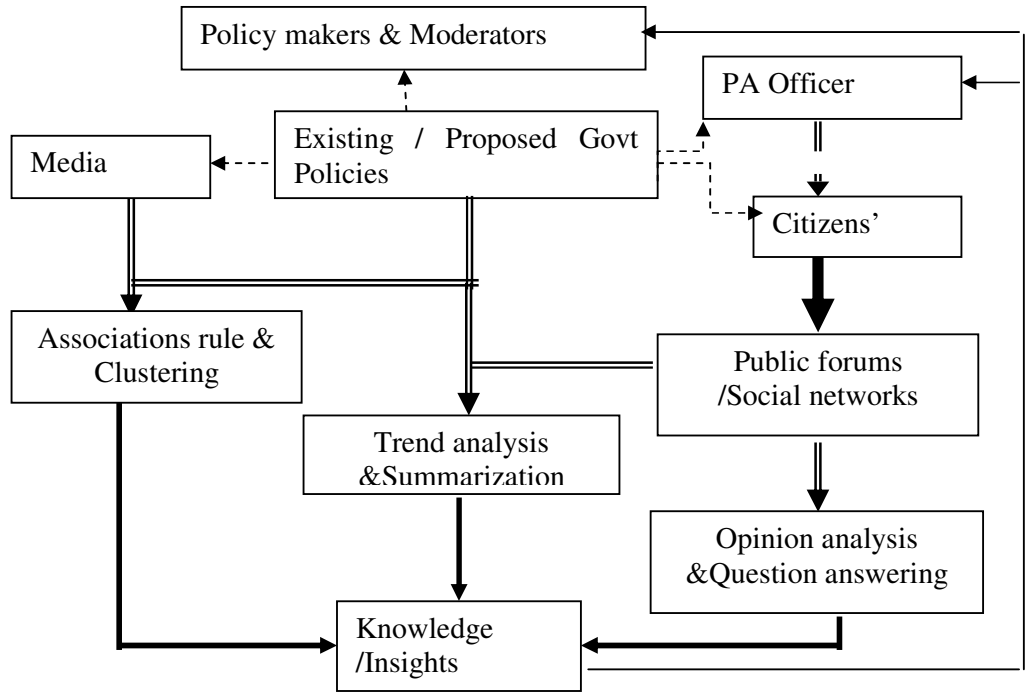

Figure. 2. Citizens' and Stakeholders' participation system

## 5. E-governance & E-democracy Projects in India

India is a land of diversity. This diversity spans across culture, language, geography and the economic condition of the people. There are significant numbers of people who are below the minimal socio-economic benchmarks. This section of the Indian society is not only deprived of basic necessities but also lack skills and elementary education. Their social development is far worse i.e. health, education, sanitation and availability of drinking water. The quality of life of these people is far below satisfactory levels thereby making the task of improving their standard of living and sustain the same is daunting. Government of India recognizes that e-governance, in the context of developing countries, provides an excellent opportunity for improving the quality of life of these sections of society and moreover it could actually provide them more equitable access to economic opportunities. India's experience in e-governance initiatives has demonstrated significant success in improving accessibility, cutting down costs, reducing corruption and increased access to un-served groups ([31],[45]). The study points out that the development of infrastructure is very important in countries such as India, which have a high proportion of global population and could benefit from E-Government if literacy can be improved [46].

E-governance is reforming the way government manages and shares information with external and internal clients. Specifically, it harnesses information and communications technologies (such as Wide Area Networks, the Internet, and mobile computing) to transform relations with citizens, businesses and amongst various arms of government. Kanungo has discussed the need of Citizen Centric e-Governance in India and discussed about the need to create a culture of maintaining, processing and retrieving the information through an electronic system and use that information for decision making [30]. The Government of India, in various forums, has indicated its commitment to provide efficient and transparent government to all strata of society. E-Governance is now mainly seen as a key element of the country's governance and administrative reform agenda. The Government of India aspires to provide [44]:

- Governance that is easily understood by and accountable to the citizens, open to democratic involvement and scrutiny (an open and transparent government)
- Citizen-centric governance that will cover all of its services and respect everyone as individuals by providing personalized services.
- An effective government that delivers maximum value for taxpayers' money (quick and efficient services).

However, E-governance is more than just streamlining processes and improving services. It's about transforming governments and renovating the way citizens participate in democracy. Misra, has discussed about the need of Citizen-centric & Criteria-based systems and Involving People in Developing Agenda for Good Governance by receiving citizens' voice. The lack of citizen-centricity in e-government acts as a 'brake' in the faster growth of internet penetration in India [29].

## 6. ROAD MAP FOR TEXT MINING BASED DSS IN INDIA

E-Government can advance the agenda on Governance and fiscal reform, transparency, anti-corruption, empowerment and poverty reduction .E-Governance in India has steadily evolved from computerization of Government Departments to initiatives that encapsulate the finer points of Governance, such as citizen centricity, service orientation and transparency. Paramjeet Walia (2009) has discussed about the initiative applications of Information and Communication Technologies (ICTs) in support of e-government initiatives in India [31], National portal of India is initiated as a Mission Mode Project under the National e-governance Plan (NeGP) [32] and other planning initiatives undertaken by the Government of India (GOI) have discussed about the importance of feedback pertaining to utility of the projects, which are part of NeGP (Figure 3 ) and need of a systems to assess the usefulness and impact of e-governance initiatives in India. The plan envisages creation of right environments to implement Government to Government (G2G), Government to Business (G2B), Government to Employee (G2E), and Government to Citizen. Among national portals in the Southern Asia region, India has the highest ranking portal with the highest online services score. It has the most e-services and tools for citizen engagement in the region but not included one among the top 20 countries in e-participation (United Nations E-Government Survey 2010) [33], there is not much literature available on this. Indian government should take the initiative to encourage citizens to send their feedback, complaints, and suggestions through e-portal and discuss various issues on government services in virtual discussion forums.

Gupta, has discussed about the problems with existing systems and implemented an Indian Police Information System and that can be used to extract useful information from the vast crime database maintained by National Crime Record Bureau (NCRB) and find crime hot spots using crime data mining techniques such as clustering etc. [37]. Choudhury, has noted many e-government projects which are running in India (Rural and urban level projects, National level, state level, district level projects and so on) all these projects are taking about G2C and few of them are G2G [38] and we can find very few efforts towards C2G (e-democracy).Monga has discussed about the need of making policy based on computerization to overcome environmental changes and need of series of efforts to achieve this.  Need of establishing complete connectivity between various ministries and departments so that transfer of files and papers could be done through Internet thereby choosing efficacious speed as an alternative to manual labour [39]. IIMs are working on Impact assessment of e-government projects, how e-government helps public sector to improve its performance, Critical success factors for individual projects etc.

Lata et al (2006), have discussed the major challenges and bottlenecks for successful E-governance Implementation in India. It has been shown that lack of local language interface is a major detrimental effect for wider proliferation of E-Governance applications in India. For successful deployment of E-Governance applications in multilingual domain, various standardization aspects related to input mechanisms, storage and retrieval, and output and display mechanism need to be addressed in a national perspective. It is also necessary that open-standards to be in place and adopted for seamless access and interchange information and Moreover, various research aspects for futuristic tools such as Cross-Lingual Information Retrieval between Indian Languages and W3C compliant Indian Language Web-Browsers need to be initiated in an urgent basis [54].

## 6.1. Multilingual Text Mining (MLTM)

Ralf Steinberger , has discussed about the need for highly multilingual text mining applications (10, 20 or more languages), but the available systems cover only few number of languages and also noted that machine learning solutions are particularly promising to achieve high multilingualism. Multilingual text processing is useful because the information content found in different languages is complementary, both regarding facts and opinions [47]. Hsin-Chang Yang *et al* (2010) have proposed a text mining method to extract associations between multilingual texts and use them in multilingual information retrieval. Documents written in different languages were first clustered and organized into hierarchies using the growing hierarchical self-organizing map model. They have also noted that in the domain of multilingual text mining, little attention has to be paid for building multilingual document hierarchies and deriving associations from such hierarchies [48]. Rowena Chau et al (2004), have discussed about the multilingual text mining approach to cross-lingual text retrieval (CLTR), and their multilingual text mining approach for automatically discovering the multilingual linguistic knowledge contributes to cross-lingual text retrieval by providing a more affordable alternative to the costly manually constructed linguistic resources. By exploiting a parallel corpus covering multiple languages, the automatic construction of language-independent concept space capturing all conceptual relationships among multilingual terms is accomplished [49].

### 6.1.1.Multilingual and Cross Lingual Projects in India

India is a multi-lingual with 22 official languages (Table 2) and multi-script (Fig. 4) country. The Indian languages belong to four language families namely Indo-European, Dravidian, Austro-Asiatic (Austric) and Sino-Tibetan. Majority of India's population are using Indo-European and Dravidian languages. The former are spoken mainly in northern and central regions and the latter in southern India. Some ethnic groups in Assam and other parts of eastern India speak Austric languages. People in the northern Himalayan region and near the Burmese border speak Sino-Tibetan languages. As the amount of textual data on the Internet increases, there are also an increasing number of people who want to retrieve information in their native language. Many citizens also have multilingual capabilities that allow them to understand more than one language [34]. This is one of the main reasons behind developing cross-language information retrieval systems. It is therefore essential that tools for information processing in local languages are developed in India. Development of technologies in multilingual computing areas involves intensive indigenous R&D efforts due to variety of Indian languages. The focused areas of the Technology Development for Indian Languages Programme in India may be divided into following domains [43]:

-- Translation Systems           - Cross Lingual Information Access and Retrieval
-- Linguistic Resources              -- Human Machine Interface systems
-- Language processing and Web tools     -- Localization and content creation

Figure . 4. Some major Indian Language scripts [34]

The CLIA (Cross Lingual Information Access) Project is a mission mode project funded by Government of India; it is an extension of the Cross-Language Information Retrieval paradigm (CLIR) ([34-36]). Cross-Language Information Access exploits the advantage of multilingual capability of users and expands search bandwidth by providing the content which is available in other language also. By using CLIR users can give queries in their native language and retrieve documents, whether in the same language as the query is, are relevant documents are found in any other language. The main components in our cross lingual information retrieval system are i) Language Analyzer, ii) Named Entity recognizer, iii) Query Translation engine, iv) Query Expansion and v) Ranking. Cross-Language Information Access (CLIA) is an extension of the Cross-Language Information Retrieval paradigm. Users who are unfamiliar with the language of documents retrieved are often unable to obtain relevant information from these documents. The objective of CLIA is to introduce additional post retrieval processing to enable users make sense of these retrieved documents.

### 6.1.2 Machine Translation and CLIA Achievements during 2010-11

Machine Translation (MT)( English to Indian Language):  In the Phase-1 of the project English-Indian Languages Machine Translation Systems (EILMT) (Figure.5) for 8 Language Pairs: English to Hindi, Marathi, Bengali, Oriya, Tamil, Urdu, Punjabi and Malayalam in the tourism domain with varying efficiency have been completed. . The project is funded by Department of Information Technology, MCIT, and Government of India. The project started from September 2006. Consortium Members of EILMT system are listed in Table 3[51].

| C-DAC MUMBAI | IISC BANGALORE |
|---|---|
| IIT HYDERABAD | C-DAC PUNE |
| IIT MUMBAI | JADAVPUR UNIVERSITY, KOLKATA |
| IIIT ALLAHABAD | UTKAL UNIVERSITY ,BANGALORE |
| AMRITA UNIVERSITY ,COIMBATORE | BANASTHALI VIDYAPEETH, BANASTHALI |

**Table 3**

Indian Language to Indian Language: Machine Translation Systems for 9 Bidirectional Language Pairs: Telugu-Hindi, Hindi-Tamil, Urdu-Hindi, Kannada-Hindi, Punjabi-Hindi, Marathi-Hindi, Bengali-Hindi, Tamil-Telugu, Malayalam-Tamil with varying efficiency have been developed. The Phase II of these projects is being initiated to improve the technology and to extend the technology to more language pairs and domains [51].

Cross-Lingual Information Access system (CLIA): In Phase-I CLIA a system for 6 Languages: Hindi, Bengali, Tamil, Marathi, Telugu and Punjabi, have been developed for the tourism domain. Under Phase II other domain will be explored [51]. Figure 6 helps us to understand the information access from English to other Indian languages.

Figure 6. Cross Lingual information access integrated with Machine Translation

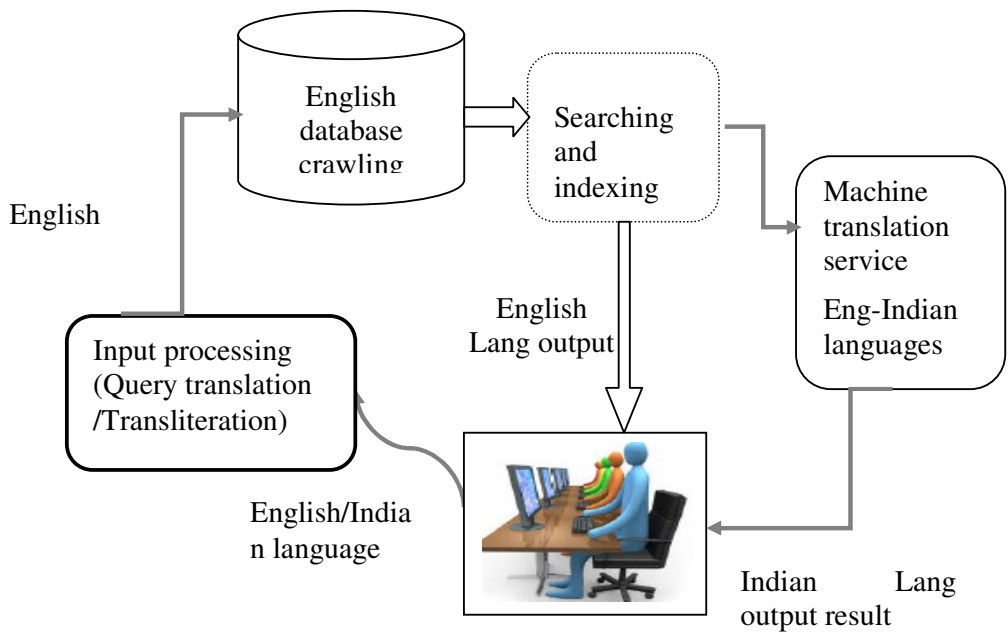

### 6.1.3 Example of Tamil-English CLIR system

The Forum for Information Retrieval Evaluation (FIRE) is an initiative in this direction [55].

The main objectives of FIRE are

   i. To encourage research in Indian language Information Access technologies by providing reusable large-scale test collections for Indian language information retrieval (ILIR) experiments

   ii. To provide a common evaluation infrastructure for comparing the performance of different IR system

   iii. To investigate evaluation methods for Information Access techniques and methods for constructing a reusable large-scale data set for ILIR experiment

R.K Rao et al, have worked on Tamil-English cross lingual information retrieval system used in the FIRE ad-hoc evaluation task. Their approach is based on bilingual dictionaries and ontology. The use of ontology for query expansion gives a significant increase in the recall without disturbing the precision. They have found that the system performs well for queries for which the word knowledge has been imparted [55].

Gyan Nidhi: Multi-Lingual Aligned Parallel Corpus consists of text in English and 12 Indian languages. It aims to digitize 1 million pages altogether containing at least 50,000 pages in each Indian language and English. Vishleshika is a tool for Statistical Text Analysis for Hindi extendible to other Indian Language texts, it examines input text and generates various statistics, e.g.: Sentence statistics, Word statistics and Character statistics [41]. Karunesh Arora et al (2004), have discussed the process for automatic extraction of phonetically rich sentences from a large text corpus for Indian languages. The importance of such a system and an algorithm to generate a set of phonetically rich sentences from a large text corpus is described along with the results for Hindi language [42]. C-DAC and other R&D organizations' are working on various projects related to Multilingual Information retrieval, Data Mining, statistics, machine learning and natural language processing projects.

**Table 4.** Institues which are working e-govt projects assesment,NLP,and CLIR

| Institute | Institute | Institute |
|---|---|---|
| IIT Madras | AU-CEG Chennai | ISI Kolkata [ISI] |
| IIIT Hyderabad | AU-KBC Chennai | Jadavpur University [JU] |
| IIT Bombay | CDAC Noida [CDACN | Utkal University [UU] |
| IIT Kharagpur | CDAC Pune [CDACP] | CDAC  Mumbai |
| CDAC Bangalore | IIM Ahmadabad | MIT and IIIT- Bangalore |

### 6.1.4 Text Mining based DSS for Tourism of Orissa

Suhag sundar et al, have implemented a tourist decision support system that mines data regarding tourist places in Orissa from Oriya text files, translates and pre-processes data and classifies the tourist places into three classes. The result obtained is then used to help international tourists in selecting places to visit based on their preference including locations on which very little data is available on the Internet [50].

## 6.2  Steps for TMbDSS in India

From the available literature, currently running e-government and e-democracy projects in R&D Institutes of Indian government and annual report of 2009-2010 from Department of Information Technology India [43], we can conclude that efforts devoted towards Text mining based citizen-centric solutions was limited. Text mining based DSS implementation needs a centralized initiative but decentralized implementation framework.  By examining currently running ICT projects in India and technologies used in those projects such as CLIR, Text analysis, NLP, Machine Learning, Data Mining, and Text mining in tourism and Multi-lingual Information retrieval, one can conclude that India has enough technical experts and domain expertise to start a Text mining initiative. The way forward would be:

— Do a detailed study to find the ways and create a strategic plan
— Bring people from Institutes like IITs, ISI, IIMs ,IIITs,C-DAC etc and form an association
    o   IITs, IIITs and ISI,AU,JU,UU etc ,can work on core part of the project
    o   C-DAC , MIT and IIIT Bangalore and IIMs can work as a bridge between R&Ds, Govt and Industry
— Start with an implementation of a pilot project at national level and replicate it to the states

All the national government documents are either in English or Hindi, So India could start a Bi-lingual (Hindi and English) TMbDSS project by using the following sample architecture (Figure 6) and then extend the same to the other Indian languages as Multi-lingual Text Mining based Decision support system (MLTMbDSS), in the future.

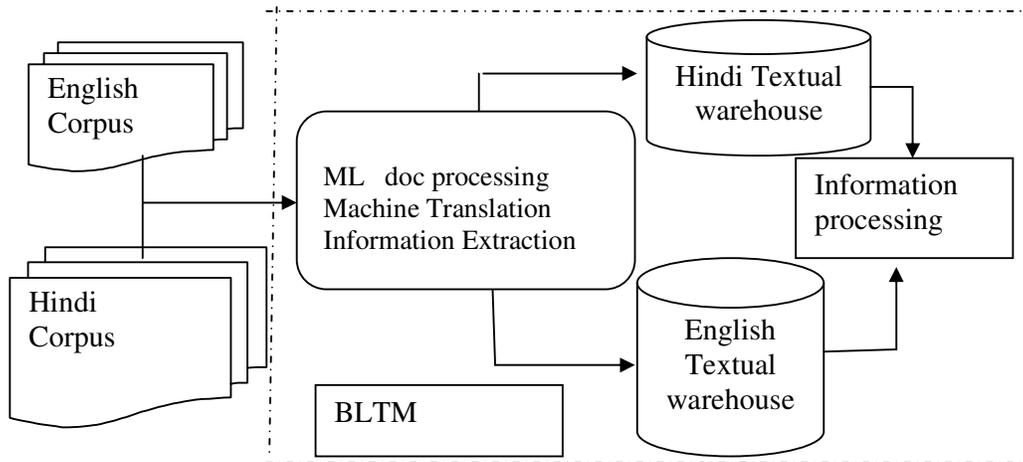

Figure 7. Technical Architecture of Bi-Lingual Text Mining Project for India

English Corpus pre-processing step involves word segmentation; stopword elimination, stemming, and keyword selection, and extracting representative keywords from a document. After these processing steps a set of keywords will be obtained that would be representative of each document. All keywords of all documents will then be collected to build a vocabulary for English keywords. Similarly Hindi documents can be processed to build a vocabulary for Hindi keywords (these documents may require different type of pre-processing steps). Each document can be encoded into a vector. Text mining techniques can then be applied to the document vector model, in the usual manner.

## 7. CONCLUSION

In this paper we have discussed need of text mining based DSS for government agencies, various text mining applications developed in e-government, architecture for system development process and proposed an integrated framework that can be used by government organizations' to develop text mining based DSS. We have also studied e-government objectives and the need for citizen-centric systems for India and provided a road map for an Indian TMbDSS project. India can start with bilingual text mining project at national level and extend the same as multi lingual text mining initiative and then replicate the system to states at a later stage.

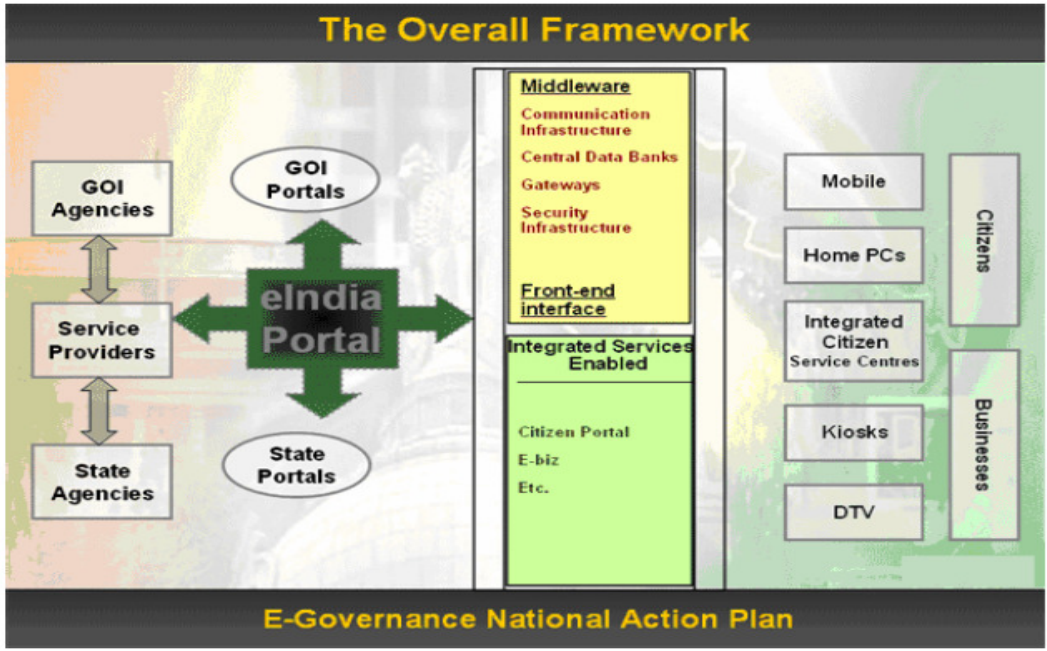

**Table 2**

| Languages | Official Language of states | Spoken by |
|---|---|---|
| Assamese | Assam | 15 million |
| Bengali | Tripura & West Bengal | 67 million |
| Bodo | Assam | |
| Dogri | Jammu and Kashmir | |
| Gujarati | Dadra and Nagar Haeli, Daman and Diu & Gujrat | 43 million |
| Hindi | Andaman and Nicobar Islands, Bihar, Chandigarh, Chhattisgarh, Delhi, Haryana, Himachal Pradesh, Jharkhand, Madhya Pradesh, Rajasthan, Uttar Pradesh & Uttaranchal | 180 million |
| Kannada | Karnataka | 35 million |
| Kashmiri | | |
| Konkani | Goa | |
| Maithili | Bihar | 22 million |
| Malayalam | Kerala & Lakshadweep | 34 million |
| Manipuri (Meithei) | Manipur | |
| Marathi | Maharashtra | 65 million |
| Nepali | Sikkim | |
| Oriya | Orissa | 30 million |
| Punjabi | Punjab | 26 million |
| Sanskrit | | |
| Santhali | | |
| Sindhi | | |
| Tamil | Tamil Nadu & Pondicherry | 66 million |
| Telugu | Andhra Pradesh | 70 million |
| Urdu | Jammu and Kashmir | 46 million |

# Authors

**Mr. G. Koteswara Rao**[1] received B.Sc degree in Mathematics, Statistics and Computer science and M.Sc in Mathematics from **IIT KANPUR**. He had worked with HCL as a Research Engineer from Oct-2007 to Oct- 2009. Currently he is working in **IIM Indore** as a Research Associate in Information Systems and Information Technology area. He has presented papers in various conferences & seminars and published papers in Springer CC&IT series, RBTR and other conference proceedings, in the areas like, E-governance, E-democracy,

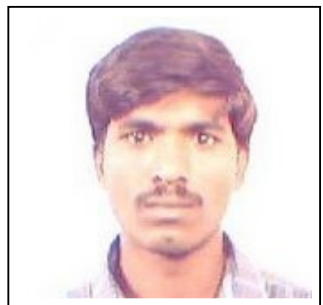

Banking Sector, Business Intelligence, Knowledge Management and Text Mining. He has also reviewed papers for ACITY-2011, CCSEIT-2011 and AOM Annual Meeting-2011.

**Prof. Shubhamoy Dey**[2] is a faculty in the area of Information Systems at IIM Indore since 2002. He has obtained his Ph.D in Data Mining and Knowledge Discovery in Databases from the School of computing, University of Leeds, U.K. He also holds B.E. and M.Tech degrees from Jadavpur University and IIT Kharagpur respectively.

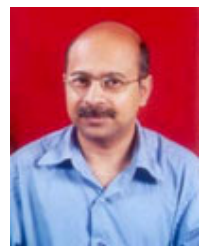

Prof. Dey`s research interests are Data mining and knowledge discovery in databases, Spatial databases, Data warehousing, Database systems, Empirical modeling and Computational finance.

He has published papers in national and international forums on Data Mining, Spatial Data Mining, Text Mining and Computational Finance. His teaching interests are Management Information Systems, Decision Support Systems, Database Systems, Enterprise Systems, Data Warehousing, Data Mining and Text & Blog Mining.

Prof. Dey carries with him rich industry experience from Hindustan cables, Wipro Information Technology, CMC Ltd., BRI (Europe), British American Consulting Group and Bank of Scotland. He has worked over 11 years in the Information Technology industry in UK and USA, and 4 years in India.

Since 1997, He has been running his own consultancy company in UK and has been providing his services as an Independent IT Consultant to major commercial organisations in UK. His consultancy clients include: Paradeep Port Trust, Berger Paints, Indian Oil Corporation, Government of Bihar, Government of Madhya Pradesh, State Bank of India, Eastern Coalfields and Department of Electronics in India; London Underground, The British Library, Fujitsu-ICL (UK), Manufacturing Science & Finance, Barretts Group Plc., Kingston Communications Group Plc., Cerillion Technologies and Barclays Bank Plc. in UK; American Stores Corporation and ALH Group Inc. in USA.